\documentclass[prb,aps,showpacs,unsortedaddress,superscriptaddress,amsmath,amssymb,twocolumn,floatfix]{revtex4}
\usepackage{graphicx}
\usepackage{amssymb}
\usepackage[usenames]{color}
\usepackage{amsmath}
\usepackage{amsthm}

\begin{document}

\title{Absence of a quantum limit to charge diffusion in bad metals}
\author{Nandan Pakhira}
\email{npakhira@gmail.com}
\author{Ross H. McKenzie}
\email{r.mckenzie@uq.edu.au}
\homepage{condensedconcepts.blogspot.com}
\affiliation{School of Mathematics and Physics, The University of Queensland, Brisbane, QLD 4072, Australia.}
\begin{abstract}
Good metals are characterised by diffusive transport of coherent quasi-particle states and the resistivity is much less than the Mott-Ioffe-Regel 
(MIR) limit, $\frac{ha}{e^{2}}$, where $a$ is the lattice constant. In bad metals, such as many strongly correlated electron materials, the resistivity  
exceeds the Mott-Ioffe-Regel limit and the transport is incoherent in nature. Hartnoll, loosely motivated by holographic duality (AdS/CFT correspondence) 
in string theory, recently proposed a lower bound to the charge diffusion constant, $D \gtrsim \hbar v_{F}^{2}/(k_{B}T)$, in the incoherent regime of transport, where $v_F$ is the Fermi velocity and
$T$ the temperature.
Using dynamical mean field theory (DMFT) we calculate the charge diffusion constant in a single band Hubbard model at half filling.
We show that in the strongly correlated regime the Hartnoll's bound is violated in the crossover region between the coherent Fermi liquid region and the 
incoherent (bad metal) local moment region. The violation occurs even when the 
bare Fermi velocity $v_F$ is replaced by its low temperature renormalised value, $v_F^*$.The bound is satisfied at all temperatures in the weakly and moderately correlated systems as well as in strongly correlated systems in 
the high temperature region where the resistivity is close to linear in temperature.
Our calculated charge diffusion constant, in the incoherent regime of transport, 
also strongly violates a proposed quantum limit of spin diffusion, $D_{s} \sim 1.3 \hbar/m$, where $m$ is the fermion mass, 
experimentally observed and theoretically calculated in a cold degenerate Fermi gas in 
the unitary limit of scattering.    
\end{abstract}
\maketitle

\section{Introduction}
Good metals like copper and gold are characterised by high optical reflectivity, electrical and thermal conductivity. The transport in these systems can be 
characterised by diffusive transport of coherent quasi-particle states, where the mean-free path is much larger than the lattice constant. The low temperature 
resistivity in good metals is well within the Mott-Ioffe-Regel (MIR) limit, $\frac{ha}{e^{2}}\sim 250 \; \mu\Omega-\textrm{cm}$, where $a$ is the lattice constant.
However, in a large class of strongly correlated systems like $3d$-transition metal oxide compounds and most notably in the strange metal regime of doped cuprates 
(high $T_{c}$ superconductors) at optimal doping the resistivity far exceeds the MIR limit~\cite{HusseyPhilMag2004} and hence cannot be characterised by diffusive 
transport of coherent quasi-particle states in the limit of weak scattering. Other signatures of a bad metal include a thermopower of order $k_{B}/e$, the absence 
of a Drude peak in the optical conductivity, and a non-monotonic temperature dependence of the Hall constant and thermopower.~\cite{MerinoPRB2000,DengPRL2013,XuPRL2013}

There have been a range of theoretical attempts to understand the incoherent regime of transport, especially for the strange metal phase of doped cuprates (high $T_{c}$ 
superconductors) at optimal doping. Recently, there is a string theory based approach to understand transport in the incoherent 
regime~\cite{SachdevJPcond2009,FaulknerScience2010}. String theory, originally proposed as a possible theory for quantum gravity, is mathematically consistent but yet has 
no experimental verification. In the following paragraph we briefly describe how a string theory based approach has been proposed to describe transport in condensed matter 
systems.       
 
Maldacena conjectured~\cite{Maldacena1998} that the large $N$ limit of certain supersymmetric conformal field theory (CFT) has correspondence to super gravity 
in anti-de Sitter spaces in higher dimension. This is known as the \textit{AdS/CFT correspondence} or \textit{gauge/gravity duality}. The most famous example 
of AdS/CFT correspondence states that IIB string theory in the product space $AdS_{5}\times S^{5}$ is dual to large $N_{c}$ limit of $\mathcal{N}=4$ 
supersymmetric $SU(N_{c})$ Yang-Mills theory on the four dimensional boundary. Further AdS/CFT correspondences relate fluid dynamics to event horizon dynamics 
of a black hole in anti-de Sitter space. In the hydrodynamic regime (long wavelength limit) of the correspondence, Einstein's equations of general relativity 
reduce to the Navier-Stokes equation for fluid mechanics. Classical fluids are characterised by transport coefficients such as shear viscosity and diffusion constant. 
Using the AdS/CFT correspondence Kovtun \textit{et al.}~\cite{KovtunPRL2005} calculated the ratio, $\eta/s$, of the shear viscosity ($\eta$) and the entropy density 
($s$) and proposed a lower bound $\frac{\eta}{s}\geq \frac{\hbar}{4\pi k_{B}}$. Such a bound is respected in classical fluids like water, the quark-gluon plasma (QGP) 
created in the relativistic heavy ion collider (RHIC)~\cite{Shuryak2004}, and in experiments on cold degenerate Fermi gases in the unitary limit~\cite{CaoScience2011}.
 However, some violations of this bound have been reported \cite{Schafer}.
Inspired by the bound on the viscosity and also
using the AdS/CFT correspondence Hartnoll recently proposed~\cite{Hartnoll2014} a lower bound for the diffusion constant, 
\begin{eqnarray}
D\gtrsim \mathcal{D}_{H} \equiv \frac{\hbar v_{F}^{2}}{k_{B} T},
\end{eqnarray}
in the incoherent regime of transport in strongly correlated electron systems. But, except near quantum critical point, condensed matter systems are probably neither 
relativistic nor conformal~\cite{AndersonPT2013}. So, this proposal needs to be tested against model based calculations.      

The temperature dependent diffusion constant, $D(T)$, is related to the temperature dependent conductivity, 
$\sigma(T)$, through the \textit{Nernst-Einstein relation}
\begin{eqnarray}
\label{Eq:NernstEinstein}
\sigma(T)=e^{2}\frac{\partial n}{\partial \mu} D(T).
\end{eqnarray}
where $\kappa_{e}(T) = \frac{\partial n}{\partial \mu}$ is the \textit{charge compressibility}. For completeness we give a derivation of this relation in the Appendix. 
Because of the above relation knowledge of the charge diffusion constant, $D(T)$, may
help us to better understand the electrical conductivity, $\sigma(T)$. 

Experiments on cold degenerate Fermi gases in the unitary scattering
limit show a quantum limit to the spin diffusion constant~\cite{SommerNature2011,BardonScience2014}, 
$D_{s}\simeq 1.3 \hbar/m$. This bound is also supported by theoretical calculations \cite{EnssPRL2012}. However, experiments on a  two-dimensional 
Fermi gas found a value $D_{s}$ that was more than two orders of magnitude smaller than the proposed bound~\cite{KoschorreckNaturePhys2013}. But spin diffusion in charge neutral 
systems such as cold atomic gases has no obvious relation to charge diffusion in charged quantum fluids such as strongly correlated electron systems. For example, in a Mott 
insulator the charge diffusion constant is zero but the spin diffusion constant is non-zero. Hence, it is conceivable that in a bad metallic phase close to the Mott insulator 
that the charge diffusion constant is much smaller than the spin diffusion constant. In the present article we do a model based calculation of the charge diffusion constant in a 
Hubbard model, and explore the possible existence of a lower bound to the charge diffusion constant and its possible relation to spin diffusion in atomic gases.

The organization of the paper is as follows. In Sec. II we introduce the single band Hubbard model and its solution under single site dynamical mean field theory 
(DMFT). We also briefly describe DMFT self-consistency using iterated perturbation theory (IPT) as a solver for the impurity problem arising under single site DMFT.
In Sec. III we briefly introduce calculation of transport and thermodynamic quantities under the single site DMFT approximation. Then in Sec. IV we show our results for 
a single band Hubbard model on the Bethe lattice at half-filling. We find clear violation of Hartnoll's proposed bound, even when the bare Fermi velocity
$v_F$ is replaced by the low-temperature renormalised velocity, $v_F^*$. 
Finally, in Sec. V we conclude and briefly consider how relaxing 
some of our assumptions may modify the results.  
% ==============================================================================================================================================================
\section{Model based calculations}
We consider the single band Hubbard model with nearest neighbor hopping, described by the Hamiltonian
\begin{eqnarray}
H=-t\sum_{\langle i,j\rangle, \sigma} (c_{i\sigma}^{\dagger}c_{j\sigma}+ H. c.)-\mu\sum_{i,\sigma}n_{i\sigma}+U\sum_{i}n_{i\uparrow}n_{i\downarrow}
\end{eqnarray}
where $n_{i\sigma}=c_{i\sigma}^{\dagger}c_{i\sigma}$, $t$ is the hopping amplitude $\mu$ is the chemical potential and $U$ is Coulomb repulsion for a doubly occupied site. 
This is probably the simplest model which incorporates nontrivial strong correlation effects. But this model has an exact solution only in one dimension and study of this model 
in higher dimension involves various approximations. Static mean field descriptions like the  Hartree-Fock decomposition of the quartic term $U n_{i\uparrow} n_{i\downarrow} 
\simeq U \langle n_{i\uparrow} \rangle n_{i\downarrow}+U n_{i\uparrow} \langle n_{i\downarrow} \rangle$ only shifts the local chemical potential. Because of the complete neglect 
of the quantum fluctuations this approximation does not generate any new energy scale (e.g. the Fermi liquid coherence scale) which can become relevant at low temperature regions.
However, as in the case of classical mean field theory for the Ising model, in the limit of of large dimension, $d\rightarrow\infty$, (or large connectivity $z$) the 
model reduces to an effective single impurity model provided the scaling $t\rightarrow t^{*}/\sqrt{2d}$ is made on a $d$-dimensional hyper-cubic lattice~\cite{MetznerPRL1989}. 
Under this approximation we neglect all spatial fluctuations yet fully retain quantum dynamics for the single site. The self-energy $\Sigma(\omega)$ only depends on 
frequency and not wave vector. This is known as the dynamical mean field theory~\cite{GeorgesRMP1995} (DMFT). It has been found DMFT gives a good description of the
Mott metal-insulator transition with increasing correlation strength, $U$, and the crossover from a Fermi liquid to bad metal with increasing temperature~\cite{MerinoPRB2000}.
Furthermore, DMFT has been found to give a quantitative description of the temperature dependence of the resistivity~\cite{LimelettePRL2003} and the frequency dependent optical 
conductivity~\cite{MerinoPRL2008} for organic charge transfer salts that are described by a two dimensional Hubbard model at half-filling~\cite{PowellRepProgPhys2011}.
Combining DMFT with electronic structure calculations based on density functional theory has given an excellent description of properties of a diverse range of transition 
metal and rare earth compounds~\cite{KotliarRMP2006}.    
% --------------------------------------------------------------------------------------------------------------------------------------------------------------
\subsection{Dynamical mean field theory}
As a consequence of the scaling, $t\rightarrow t^{*}/\sqrt{2d}$, all the self energy diagrams, arising under skeletal graph expansion of the irreducible self 
energy and involving non local Green's functions vanishes in the limit $d\rightarrow\infty$. Then the self energy becomes local and involves only the local Green's function. 
The lattice problem for the Hubbard model then can be mapped onto an effective single impurity Anderson model~\cite{GeorgesRMP1995} :
\begin{eqnarray}
H_{imp} &=& \sum_{l,\sigma} (\tilde{\epsilon}_{l}-\mu) c_{l\sigma}^{\dagger}c_{l\sigma}+\sum_{l,\sigma}(V_{l}c_{l\sigma}^{\dagger}d_{0\sigma}+H.c.)\nonumber \\
&&-\mu\sum_{\sigma}n_{d0\sigma}+Un_{d0\uparrow}n_{d0\downarrow},
\end{eqnarray}   
where $n_{d0\sigma}=d_{0\sigma}^{\dagger}d_{0\sigma}$. The operators $d_{0\sigma}^{\dagger}$ and $d_{0\sigma}$ characterise a given site $i=0$ and 
$\{c_{l\sigma}^{\dagger},c_{l\sigma}\}$ characterise the effective bath arising from electrons at all other sites. $\tilde{\epsilon}_{l}$ and $V_{l}$ are effective 
parameters characterising the dispersion of the bath and its coupling to the local site. $\tilde{\epsilon}_{l}$ and $V_{l}$ or equivalently the bath Green's function,
given by 
$G_{0}(\omega)$, 
\begin{eqnarray}
&&G_{0}^{-1}(\omega)=\omega+\mu-\int_{-\infty}^{+\infty}\frac{\Delta(\epsilon)\;d\epsilon}{\omega+\mu-\epsilon}\nonumber \\ 
&&\Delta(\epsilon)=\sum_{l\sigma}V_{l}^{2}\delta(\epsilon-\tilde{\epsilon}_{l})
\end{eqnarray}
can be calculated self consistently by solving the impurity problem iteratively. The solution of the impurity problem is the toughest part and usually involves use of
numerical methods such as quantum Monte Carlo (QMC), exact diagonalization (ED), or the numerical renormalization group (NRG). 

We use iterated perturbation theory 
(IPT)~\cite{ZhangPRL1993,KajueterPRL1996} as it is easy to implement, computationally cheap, and captures the essential physics in the parameter regime we are interested in, 
$U < 0.8U_{c}$ where $U_{c}$ is the critical value of $U$ at which the Mott metal-insulator transition occurs. For example, Bulla~\cite{BullaPRL1999} showed that for the Bethe 
lattice at half-filling the results of IPT and NRG are similar except extremely close to the Mott transition. Indeed in the proximity of the Mott transition, Terletska 
\textit{et al.}~\cite{TerletskaPRL2011} found that the temperature dependent resistivity calculated from IPT was in agreement with that found by continuous time QMC (CT-QMC). 
Also, recently Arsenault \textit{et al.}~\cite{ArsenaultPRB2012} showed that for lattices with a van Hove singularity in the density of states (DOS), even in the proximity of 
the Mott transition IPT with a modified self-consistency condition matches with results from CT-QMC. So, for the single band Hubbard model results from the IPT are generic in 
nature. In the next section we review DMFT self consistency using IPT.    
% --------------------------------------------------------------------------------------------------------------------------------------------------------------
\subsection{Iterated Perturbation Theory}
The iterated perturbation theory (IPT) is a semi-analytical method. The irreducible self-energy in IPT is approximated using second order 
polarization bubble involving bath Green's function, $G_{0}(\omega)$. The self-energy under this approximation can be shown 
(using moment expansion) to smoothly interpolates between the atomic limit $t=0$ and the weak coupling limit $U\rightarrow 0$. In the following paragraph we elaborate DMFT 
self-consistency method using IPT as impurity solver. We work with real, not imaginary, frequencies and so no analytic continuation is necessary.\\ 
(i) For a given lattice density of states $N_{0}(\epsilon)$ and self energy $\Sigma(\omega)$ the \textit{local} Green's function is given by
\begin{eqnarray}
G(\omega)=\int_{-\infty}^{+\infty}\frac{N_{0}(\epsilon)d\epsilon}{\omega^{+}+\mu-\Sigma(\omega^{+})-\epsilon},
\end{eqnarray}
where $\mu$ is the \textit{local} chemical potential.\\ 
(ii) From knowledge of the \textit{local} Green's function $G_{\textrm{loc}}(\omega)$ we can calculate the \textit{bath hybridization} function, 
$\Delta(\omega)$ by using- 
\begin{eqnarray}
\Delta(\omega)=\omega^{+}+\mu-\Sigma(\omega)-G^{-1}(\omega).
\end{eqnarray}
(iii) Subsequently using \textit{bath hybridization} we can calculate the \textit{bath Green's function} as
\begin{eqnarray}
G_{0}(\omega) = \frac{1}{\omega+\tilde{\mu}_{0}-\Delta(\omega)}.
\end{eqnarray}
The parameter $\tilde{\mu}_{0} = \mu-Un$ is the bath chemical potential and it vanishes at half filling for the particle-hole symmetric case, which we consider 
in the present study.\\ 
(iv) The fully interacting Green's function can be calculated using the Dyson's equation
\begin{eqnarray}
G(\omega)=\frac{1}{G_{0}^{-1}(\omega)-\tilde{\mu}_{0}+\mu-\Sigma(\omega)}.
\end{eqnarray} 
(v)The new self-energy can be calculated following the IPT ansatz~\cite{KajueterPRL1996} as
\begin{eqnarray}
\Sigma(\omega)=Un+\frac{A\Sigma^{(2)}(\omega)}{1-B\Sigma^{(2)}(\omega)}
\label{selfipt}
\end{eqnarray}
where, 
\begin{eqnarray}
A=\frac{n(1-n)}{n_{0}(1-n_{0})}\;\; ;\hspace*{0.5cm} B=\frac{U(1-n)-\mu+\mu_{0}}{n_{0}(1-n_{0})U^{2}}
\end{eqnarray}
and
\begin{eqnarray}
\label{Eq:defn_nd}
&& n = -\frac{1}{\pi}\int_{-\infty}^{+\infty} d\omega \; n_{F}(\omega)\;\textrm{Im}[G(\omega^{+})],\\
\label{Eq:defn_nd0}
&& n_{0} = -\frac{1}{\pi}\int_{-\infty}^{+\infty}d\omega \; n_{F}(\omega)\;\textrm{Im}[G_{0}(\omega^{+})]
\end{eqnarray}
are the \textit{local} and \textit{bath} particle numbers, respectively. $\Sigma^{(2)}(\omega)$ is the self energy from second order 
perturbation theory and is given by 
\begin{eqnarray}
\Sigma^{(2)}(\omega) &=& U^{2}\!\!\int\limits_{-\infty}^{+\infty}\prod_{i=1}^{3}\left(d\epsilon_{i}\rho_{0}(\epsilon_{i})\right)
\left[\frac{n_{F}(-\epsilon_{1})n_{F}(\epsilon_{2})n_{F}(-\epsilon_{3})}{\omega+i\eta-\epsilon_{1}+\epsilon_{2}-\epsilon_{3}}\right . \nonumber \\
&&\hspace*{1cm}\left . +\frac{n_{F}(\epsilon_{1})n_{F}(-\epsilon_{2})n_{F}(\epsilon_{3})}{\omega+i\eta-\epsilon_{1}+\epsilon_{2}-\epsilon_{3}}\right]
\end{eqnarray}
where $\rho_{0}(\omega)=-\frac{1}{\pi}\textrm{Im}[G_{0}(\omega^{+})]$ and $\eta\rightarrow 0^{+}$. We iterate (i) - (v) until the desired self-consistency in self-energy 
and other physical quantities are achieved. Hence, we focus solely on the case of half-filling ($n=1$). Due to particle-hole symmetry $\mu=\frac{U}{2}$ for all $U$ and $T$. 
This speeds up computation significantly, as it is not necessary to self-consistently determine $\mu$ from Eq.~(\ref{Eq:defn_nd}) 
% --------------------------------------------------------------------------------------------------------------------------------------------------------------
\subsection{Bethe lattice}
We choose a Bethe lattice (Cayley tree) because it makes computation even faster because  the local Green's function, $G(\omega)$, has an exact 
analytical form. Particle-hole symmetry also simplifies the calculations.
The Bethe lattice produces qualitatively similar results to the hyper-cubic 
lattice\cite{GeorgesRMP1995} and lower dimensional Hubbard models \cite{MerinoPRL2008,XuPRL2013}. In the limit of 
infinite coordination number ($z\rightarrow\infty$), the density of states has \textit{semi-circular} form~\cite{economou2010green} :
\begin{eqnarray}
N_{0}(\epsilon)=\frac{2}{\pi W^{2}}\sqrt{W^{2}-\epsilon^{2}}\;\Theta(W-|\epsilon|)
\end{eqnarray}
where $\Theta(x)$ is the familiar unit step function, $W=2t^{*}$ is the half-band width and the hopping amplitude in this case is scaled as 
$t\rightarrow t^{*}/\sqrt{z}$. Most importantly the \textit{local} Green's function has the exact analytical form
\begin{eqnarray}
\label{Eq:GlocBethe}
&&G(\omega)=\frac{2}{W^{2}}\left[\zeta-\sqrt{\zeta^{2}-W^{2}}\right],\\
\label{Eq:defn_zeta}
&&\zeta(\omega)\equiv\omega+i\eta+\mu-\Sigma(\omega).
\end{eqnarray}
It can be easily verified that in this case the \textit{bath hybridization} function, $\Delta(\omega)=\frac{W^{2}}{4}G(\omega)\equiv t^{*2}G(\omega)$, is proportional to the 
\textit{local} Green's function.
% ==============================================================================================================================================================
\section{Transport properties}
Using the self-consistent self energy we can calculate various quantities like \textit{dc} conductivity, charge compressibility, and diffusivity.
% --------------------------------------------------------------------------------------------------------------------------------------------------------------
\subsection{\textit{dc} Conductivity}
In the limit of $d\rightarrow\infty$ all vertex corrections to two-body correlation functions drop out~\cite{KhuranaPRL1990} and the temperature dependent 
\textit{dc} conductivity, $\sigma(T)$, can be calculated using the simple polarization bubble as~\cite{GeorgesRMP1995,PruschkePRB1993}
\begin{eqnarray}
\sigma(T)=\frac{\pi e^{2}}{\hbar}\frac{1}{\nu}\!\int\limits_{-\infty}^{+\infty}\!d\epsilon\;\Phi_{xx}(\epsilon)\!\int\limits_{-\infty}^{+\infty}\!d\omega 
\left(-\frac{\partial n_{F}(\omega)}{\partial \omega}\right)A^{2}(\omega,\epsilon)
\end{eqnarray}
where $\nu=a^{d}$ is the volume of the unit cell of a $d$-dimensional hyper-cubic lattice with lattice constant $a$,
\begin{eqnarray}
&& A(\omega,\epsilon)=-\frac{1}{\pi}\textrm{Im}\left[\frac{1}{\omega+\mu-\Sigma(\omega)-\epsilon}\right],\\
&& n_{F}(\omega) =\frac{1}{e^{\beta\omega}+1}
\end{eqnarray}
are the spectral density and Fermi function, respectively and 
\begin{eqnarray}
\label{Eq:TrDOS}
\Phi_{xx}(\epsilon)=\frac{1}{N}\sum_{\mathbf{k}}\left(\frac{\partial \epsilon_{\mathbf{k}}}{\partial k_{x}}\right)^{2}\delta(\epsilon-\epsilon_{\mathbf{k}})
\end{eqnarray}
is the \textit{transport density of states}. $N$ is the number of lattice sites. 

Because of its tree like structure the Bethe lattice has no loop and no energy dispersion relation in $\mathbf{k}$. But, by invoking the f-sum rule it can be shown 
that~\cite{ChungPRB1998,ChattopadhyayPRB2000,ArsenaultPRB2013}
\begin{eqnarray}
\Phi_{xx}(\epsilon)=\frac{1}{3d}(W^{2}-\epsilon^{2})N_{0}(\epsilon),
\end{eqnarray}
is the correct \textit{transport density of states} in the limit of $d\rightarrow\infty$. It is interesting to mention that for a Bethe lattice with 
coordination number $z$ the connectivity $K=z-1$ while that for the hyper-cubic lattice is $2d$. So, in the limit of large coordination number we can take the 
connectivity to be equal to $2d$ and we can always do the mapping $z \leftrightarrow 2d$. 
% --------------------------------------------------------------------------------------------------------------------------------------------------------------
\subsection{Charge Compressibility}
The inverse of the charge compressibility can be interpreted as the energy cost to add or remove a particle from a system. For the 
non-interacting system ($U=0$) at zero temperature ($T=0$), $\kappa_{e}=N_{0}(E_{F})$, where $N_{0}(E_{F})$ is the density of states at the Fermi level.

In a general many-body system the local particle number is given by
\begin{eqnarray}
n=\frac{1}{\nu}\int_{-\infty}^{+\infty} d\omega \ n_{F}(\omega) \sum_{\mathbf{k}} A(\mathbf{k},\omega), 
\end{eqnarray}
where, the spectral function is
\begin{eqnarray}
A(\mathbf{k},\omega)=-\frac{1}{\pi} \textrm{Im}\left[\frac{1}{\omega+\mu-\epsilon_{\mathbf{k}}-\Sigma_{\mathbf{k}}(\omega)}\right].
\end{eqnarray}
The self-energy, $\Sigma_{\mathbf{k}}(\omega)$, in the limit of $d\rightarrow\infty$ is independent of wave vector $\mathbf{k}$ and is given by $\Sigma(\omega)$. 
Hence, differentiating with respect to $\mu$, the charge compressibility, $\kappa_{e}(T)=\frac{\partial n}{\partial \mu}$, under the DMFT approximation is given by
\begin{eqnarray}
\kappa_{e}(T) &=&\frac{1}{\pi}\textrm{Im}\int\limits_{-\infty}^{+\infty} d\omega \; n_{F}(\omega)\left(1-\frac{\partial \Sigma(\omega)}
{\partial \mu}\right)\nonumber \\
 & &\times\int\limits_{-\infty}^{+\infty}\frac{N_{0}(\epsilon)\; d\epsilon}{(\omega+\mu-\Sigma(\omega)-\epsilon)^{2}}.
\end{eqnarray}
The effect of the derivative $\frac{\partial \Sigma(\omega)}{\partial \mu}$ on the charge compressiblity of a Fermi liquid was discussed previously by 
Luttinger~\cite{LuttingerPR1960} and by Hotta and Fujimoto~\cite{HottaPRB1996}. Here it does not have a closed analytical form and we evaluate it beginning with the IPT 
expression (\ref{selfipt}).
The charge compressibility 
$\kappa_{e}(T)$ is then given by
\begin{eqnarray}
\kappa_{e}(T) = \frac{\tilde{J}+\tilde{K}}{1+U(\tilde{J}+\tilde{K})},
\label{Eqn:kappaJK}
\end{eqnarray} 
where the $\tilde{J}$ term in the denominator is associated with the
Hartree term in the self energy (\ref{selfipt}).
For the Bethe lattice
\begin{eqnarray}
\tilde{J} &=& -\frac{1}{\pi} \textrm{Im} \int\limits_{-\infty}^{+\infty} d\omega\; n_{F}(\omega)\left[2-\frac{2\zeta}{\sqrt{\zeta^{2}-1}}\right]\\
\tilde{K} &=& \frac{1}{\pi} \textrm{Im} \int\limits_{-\infty}^{+\infty} d\omega\; n_{F}(\omega)\left[2-\frac{2\zeta}{\sqrt{\zeta^{2}-1}}\right]
\frac{\partial \tilde{\Sigma}_{2}(\omega)}{\partial \mu}
\end{eqnarray}
and $\zeta(\omega)$ is given by Eq.~(\ref{Eq:defn_zeta}).
If we define $\frac{\partial \rho_{\Sigma}(\omega)}{\partial \mu}=-\frac{1}{\pi}\textrm{Im}\frac{\partial \tilde{\Sigma}_{2}(\omega)}{\partial \mu}$ then
\begin{eqnarray}
\frac{\partial \rho_{\Sigma}(\omega)}{\partial \mu}&=&\int d\epsilon_{1} d\epsilon_{2}
\left[2\frac{\partial \rho_{\mathcal{G}}(\epsilon_{1})}{\partial \mu}\rho_{\mathcal{G}}(\omega-\epsilon_{1}+\epsilon_{2})\rho_{\mathcal{G}}(\epsilon_{2})\right .\nonumber \\
&&+\left . \rho_{\mathcal{G}}(\epsilon_{1})\rho_{\mathcal{G}}(\omega-\epsilon_{1}+\epsilon_{2})\frac{\partial \rho_{\mathcal{G}}(\epsilon_{2})}{\partial \mu}
\right]\nonumber\\
&&\times\left[n_{F}(-\epsilon_{1})n_{F}(-\omega+\epsilon_{1}-\epsilon_{2})n_{F}(\epsilon_{2})\right .\nonumber \\
&&\left . +n_{F}(\epsilon_{1})n_{F}(\omega-\epsilon_{1}+\epsilon_{2})n_{F}(-\epsilon_{2})
\right],
\label{Eq:drhodmu}
\end{eqnarray}
where $\rho_{\mathcal{G}}(\omega)=-\frac{1}{\pi}\textrm{Im} G_{0}(\omega)$ and for Bethe lattice
\begin{eqnarray}
\frac{\partial \rho_{\mathcal{G}}(\omega)}{\partial \mu} = -\frac{1}{\pi}\textrm{Im}\frac{2-\frac{2\zeta}{\sqrt{\zeta^{2}-1}}}{4(\omega+\mu_{0}-\Delta(\omega))^{2}}.
\end{eqnarray}
The expression in Eq.~(\ref{Eq:drhodmu}) is calculated by using standard FFT routine and the real part of $\frac{\partial \tilde{\Sigma}_{2}(\omega)}{\partial\mu}$ 
can be calculated using Hilbert transform. 
We note in passing that we find for most parameter regimes that the
expression (\ref{Eqn:kappaJK}) is dominated by the $\tilde{J}$ terms and
the $\tilde{K}$ terms involve only a small correction.
% --------------------------------------------------------------------------------------------------------------------------------------------------------------
\subsection{Diffusivity}
As mentioned earlier the diffusivity, $D(T)$ can be calculated using the \textit{Nernst-Einstein relation} in Eq.~(\ref{Eq:NernstEinstein}).
To compare to the limit of diffusion constant, proposed by Hartnoll, we need to find the Fermi velocity $v_F$.
First, one has to decide whether this should be the bare Fermi velocity,
i.e. the “band structure” value, or a renormalised value 
$v_F^*$ associated with a low temperature Fermi liquid state.
In that case, $v_F^*=Z v_F$ where  $Z$
is the quasi-particle renormalization factor which can be  calculated from the self energy, 
$\Sigma(\omega)=\Sigma_{R}(\omega)+i\Sigma_{I}(\omega)$ :
\begin{eqnarray}
Z=\left(1-\left . \frac{\partial \Sigma_{R}(\omega)}{\partial \omega}\right|_{\omega \rightarrow 0}\right)^{-1}.
\end{eqnarray}
The Hartnoll bound, $\mathcal{D}_{H}$ then gets renormalized to $\mathcal{D}^{*}_{H}=Z^{2}\mathcal{D}_{H}$. Note that as the Mott transition is approached
this decreases the lower bound by several orders of magnitude, making it harder
to violate.
It is not completely clear to us from the arguments of Hartnoll whether one should
use $v_F$ or $v_F^*$, particularly as he is concerned with incoherent transport, i.e.,  outside
the Fermi liquid regime. Here, we use the latter but note that this choice makes the bound much less stringent.

Second, there is the issue of how to
evaluate the Fermi velocity in the DMFT approximation, in the limit of infinite 
dimensionality, $d\rightarrow\infty$. 
Since, $v_{\mathbf{k}}^{2}=\left(\frac{\partial \epsilon_{\mathbf{k}}}{\partial \mathbf{k}}\right)^{2}$ appears in the expression 
for \textit{transport density of states} in Eq.~(\ref{Eq:TrDOS}) we define 
\begin{eqnarray}
\hbar v_{F}^{2} = \frac{1}{\hbar}\frac{\Phi_{xx}(\epsilon=0)}{N_{0}(\epsilon=0)}
\end{eqnarray}
in the limit of $d\rightarrow\infty$. This definition of $\hbar v_{F}^{2}$ gives the correct Fermi velocity~\cite{PruschkePRB1993} for the hyper-cubic lattice in the 
limit of $d\rightarrow\infty$. Also, by pure dimensional analysis for any lattice structure $\hbar v_{F}=\lambda W a$ ($\lambda$ being a numerical constant of order one 
for a given lattice). Hartnoll's proposed quantum bound for diffusion constant on the Bethe lattice is then given by  
\begin{eqnarray}
\mathcal{D}_{H} = \frac{W^{2}a^{2}}{3d\hbar k_{B}T}.
\end{eqnarray}

Including the renormalisation of the Fermi velocity the dimensionless scaled diffusivity is then given by
\begin{eqnarray}
\frac{D(T)}{\mathcal{D}_{H}^{*}} &=&\pi\left(\frac{k_{B}T}{W}\right)\frac{1}{Z^{2}\tilde{\kappa}_{e}(T)}
\int\limits_{-W}^{+W} d\epsilon \int\limits_{-\infty}^{+\infty}d\omega (W^{2}-\epsilon^{2})N(\epsilon)\nonumber \\ 
&&\hspace*{2cm}\times A^{2}(\epsilon,\omega) \left(-\frac{\partial n_{F}(\omega)}{\partial \omega}\right),
\end{eqnarray}
where $\tilde{\kappa}_{e}(T)=\frac{\partial \tilde{n}}{\partial \tilde{\mu}}$ is the dimensionless charge compressibility and $\mu=\tilde{\mu}W$, 
$\tilde{n}=n\nu$. The advantage of calculating scaled diffusivity is that it does not depend on universal constants such as $\hbar$ or material dependent constants 
such as the lattice constant, $a$, and the unit cell volume, $\nu$, and the temperature appears only as a dimensionless scaled quantity.    
 
We now turn to comparison with the proposed bound for the spin diffusion constant. In a similar spirit we use $\frac{1}{m}=\frac{1}{\hbar^{2}}\frac{\partial^{2}
\epsilon_{\textrm{k}}}{\partial k_{x}^{2}}$ as a generalized definition for inverse mass. Then for the hyper-cubic lattice we get $\frac{Wa^{2}}{d\hbar^{2}}$ as an effective 
inverse mass averaged over the Fermi surface at half-filling. If we take this to be same in the Bethe lattice as well then we will have 
\begin{eqnarray}
\mathcal{D}_{A}= \frac{\alpha W a^{2}}{d\hbar}
\end{eqnarray} 
with $\alpha=1.3$ a dimensionless constant. 

In an interacting system the bare mass $m$ gets renormalized to an effective mass 
$m^{*}=m/Z$
where $Z$ is the quasi-particle weight.
Thus the bound $\mathcal{D}_{A}$ will get renormalized to 
\begin{eqnarray}
\mathcal{D}_{A}^{*} = \alpha \frac{\hbar}{m^{*}} = Z \mathcal{D}_{A}
\end{eqnarray}
The scaled diffusivity in this case is then given by
\begin{eqnarray}
\frac{D(T)}{\mathcal{D}_{A}^{*}} &=& \frac{\pi}{3\alpha Z}\frac{1}{\tilde{\kappa}_{e}(T)}\int\limits_{-W}^{+W} d\epsilon 
\int\limits_{-\infty}^{+\infty}d\omega (W^{2}-\epsilon^{2})N(\epsilon) \nonumber \\ 
&&\hspace*{2cm}\times A^{2}(\epsilon,\omega) \left(-\frac{\partial n_{F}(\omega)}{\partial \omega}\right).
\end{eqnarray}
% ========================================================================================================================================================
\section{Results}
We consider the case of half filling, $n=1$, \textit{i.e.} each site on the average is occupied by one electron. We study spectral and transport properties 
as a function of correlation strength $U$ and temperature, $T$ (enters as $k_{B}T$ with dimension of energy). Henceforth, unless stated otherwise, all the energy 
scales will be measured in units of half-bandwidth, $W$.
% --------------------------------------------------------------------------------------------------------------------------------------------------------------
\subsection{Spectral function}
\begin{center}
\begin{figure}[htbp]
\includegraphics[scale=0.32,clip=]{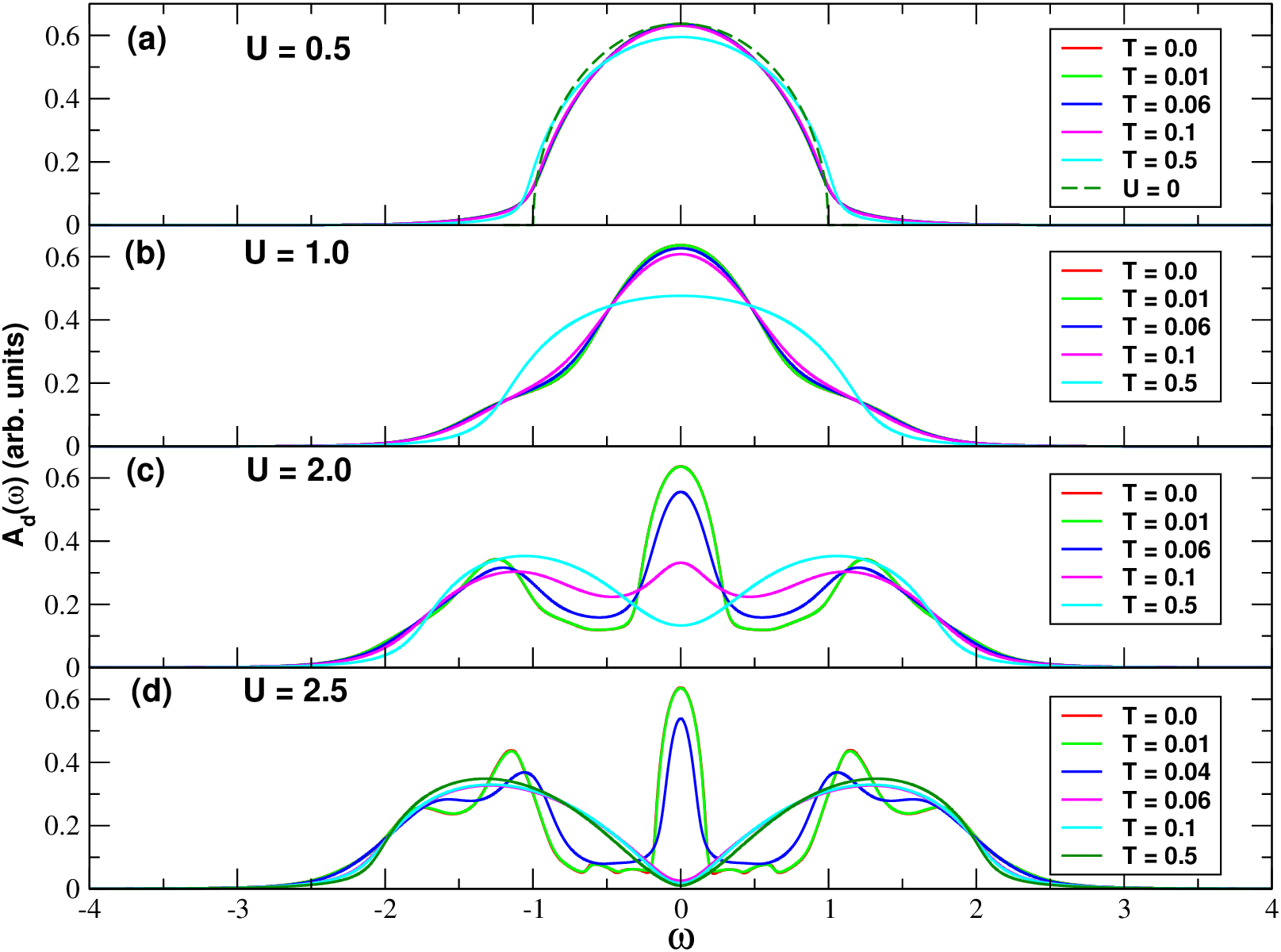}
\caption{(Color online) Energy dependent spectral function, $A_{d}(\omega)$, for various values of the interaction strength $U$ and temperature $T$. Panel (a) : weakly 
correlated regime. (Green) dashed line is the density of states for the non-interacting case. Panel (b) : moderately correlated regime. Panel (c) and (d) : strongly 
correlated regime. With increasing $U$, the integrated spectral weight under central peak (Kondo resonance) gets transferred to high energy Hubbard bands which corresponds 
to destruction of quasi-particle states. In these cases, there is a temperature dependent crossover between strong coupling regime (Fermi liquid) and local moment regime
 (bad metals). All energies and temperatures are measured in units of $W$, the half-bandwidth.}
\label{Fig:SpectralFn}
\end{figure}
\end{center}

In Fig.~\ref{Fig:SpectralFn} we show the evolution of the spectral function,
\begin{eqnarray}
A_{d}(\omega)=-\frac{1}{\pi}\textrm{Im}\left[G(\omega^{+})\right],
\end{eqnarray}
as a function of $U$ and $T$. Similar results has been obtained previously by other authors~\cite{ZhangPRL1993}. For completeness we show these results here because they 
illustrate the essential physics (the destruction of quasi-particles) behind violation of the MIR limit and Hartnoll's bound. In panel (a) of 
Fig.~\ref{Fig:SpectralFn} we show the spectral function for a weakly correlated system $U=0.5$. The spectral function is dominated by a broad central peak and a very 
small smearing of the non-interacting ($U=0$) band edges at $\omega=\pm 1$. The integrated spectral weight is dominated by the contribution from the central peak 
and $A_{d}(0)\simeq 2/\pi$, as in the non-interacting case. At finite temperature, due to particle-hole excitations across the Fermi surface, the spectral weight 
at the Fermi energy $\omega=0$ gets transfered to finite frequency but the central peak still remains intact.

As we increase the correlation strength ($U=1$) side bands develop on either side of the central peak as shown in the panel (b) of Fig.~\ref{Fig:SpectralFn}. The side 
bands eventually develop into high energy Hubbard bands at $\omega = \pm \frac{U}{2}$ as shown in panel (c) of Fig.~\ref{Fig:SpectralFn} for $U=2$. The Hubbard 
bands are well separated from the central peak which arises due to Kondo resonance effects in the effective single impurity Anderson model~\cite{GeorgesRMP1995}. The width and 
height of the Kondo resonance is controlled by the effective Kondo temperature $T_{K}$. Since, $T_{K}\sim W\exp(-\Gamma_{\textrm{eff}}/U)$, where $\Gamma_{\textrm{eff}}$ is an 
effective hybridization strength, the width of the Kondo resonance decreases while its height increases with increasing $U$ as shown in panel (d) for $U=2.5$. For very 
large $U > U_{c}$ the Kondo resonance gets completely killed and we enter into the Mott insulating state characterised by fully gapped spectral function at the Fermi energy. 
The numerical value of $U_{c}$ depends on the numerical technique that one uses and $U_{c} \simeq 3.3-3.4$ for the IPT based impurity solver~\cite{ZhangPRL1993,BullaPRL1999}.

At finite temperature in the moderately correlated regime like $U=1.0$ the central peak, despite getting broadened, remains intact even for temperatures as high 
as $T\sim W$. The side bands thermally broaden out. For the strongly correlated regime of $U=2$ and $U=2.5$ the central peak (quasi-particle peak) as well as the 
integrated spectral weight under it decreases with increasing temperature and eventually the central peak gets completely destroyed for temperatures 
$T \gg T_{K}$. This corresponds to the finite temperature crossover from the \textit{strong coupling regime} into the \textit{local moment regime} of the 
effective Anderson impurity model. The crossover region becomes increasingly sharp as evident in panel (d), which corresponds to the fragile nature of the quasi-particle state.
% --------------------------------------------------------------------------------------------------------------------------------------------------------------
% --------------------------------------------------------------------------------------------------------------------------------------------------------------
\subsection{Quasi-particle weight}
In Fig.~\ref{Fig:CompressibilityAndZT0.0} we show the continuous decrease of the quasi-particle renormalization factor, $Z$, with increasing $U$. This also tracks the 
continuous destruction of coherent quasi-particle states.Comparison of $Z$ against results from numerical renormalization group (NRG) based calculations by 
Bulla~\cite{BullaPRL1999} validates the qualitative correctness of IPT based approach though $Z$ begins to differs by 50\% in the strong correlation regime ($U=2.5$).
We also note that for $U=2.5$, $Z > 0.2$ and so in some sense for that regime the system is not extremely correlated. Yet we will see that even in this regime Hartnoll's 
bound is violated. For comparison, in the doped Hubbard model on the square lattice (with $U=16t=3.5W$ at 15\% doping $n=0.85$) DMFT gives $Z\simeq 0.2$~\cite{XuPRL2013}.
% --------------------------------------------------------------------------------------------------------------------------------------------------------------
\subsection{Charge compressibility}
\begin{center}
\begin{figure}[htbp]
\includegraphics[scale=0.36,clip=]{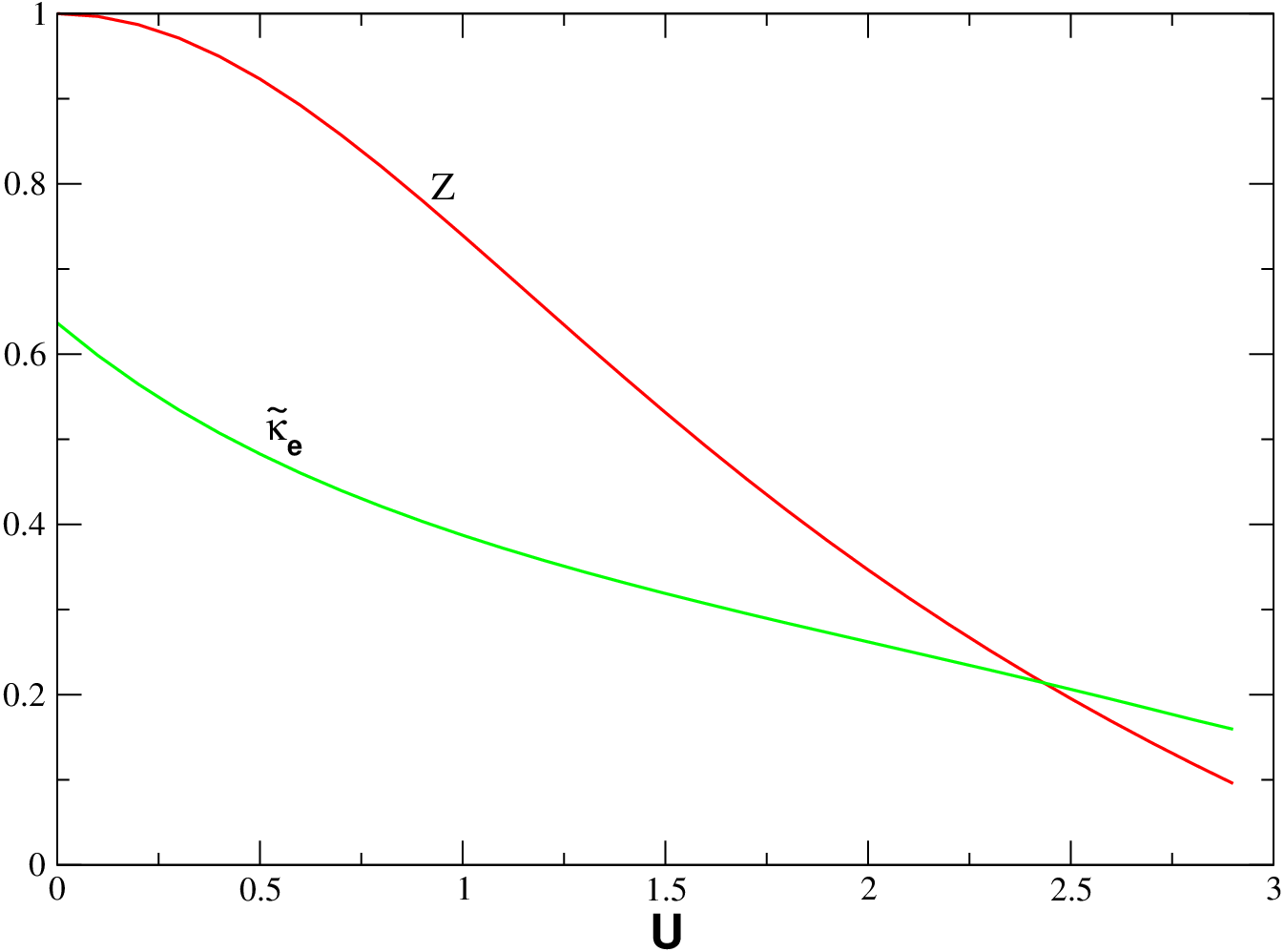}
\caption{(Color online) The zero-temperature quasi-particle renormalization factor $Z$ and charge compressibility $\kappa_{e}$ as a function of $U$. The system becomes 
increasingly incompressible with increasing $U$ and the quasi-particle weight $Z$ smoothly decreases, as the transition to the Mott insulator is approached at 
$U_{c}\simeq 3.4$ \cite{BullaPRL1999}. $U$ is measured in units of $W$.}
\label{Fig:CompressibilityAndZT0.0}
\end{figure}
\end{center}

\begin{center}
\begin{figure}[htbp]
\includegraphics[scale=0.36,clip=]{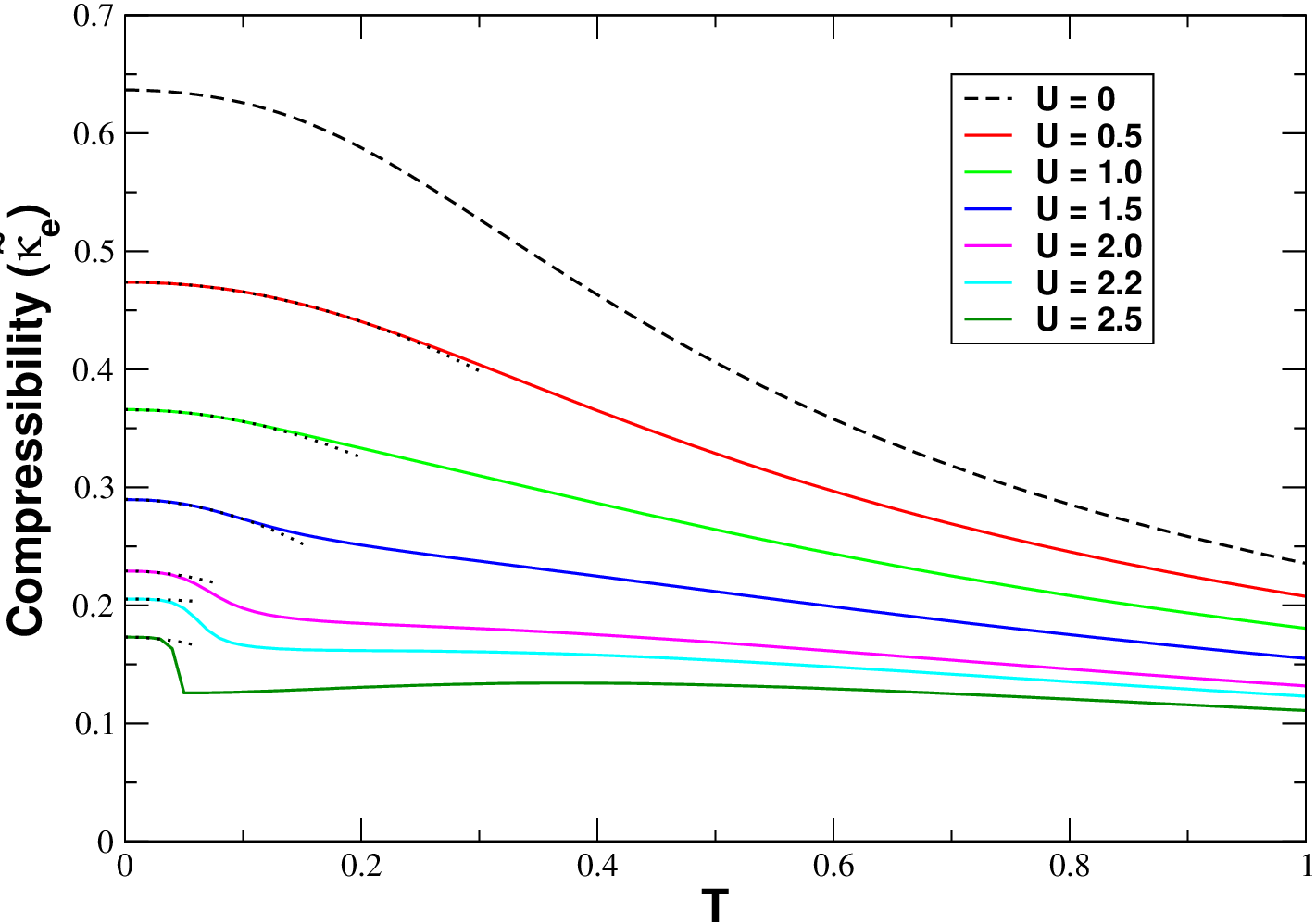}
\caption{(Color online) Temperature dependence of the charge compressibility, $\kappa_{e}$, for various correlation strengths $U$. The dashed line is the non-interacting 
($U=0$) case calculated using Eq.~(\ref{Eq:KappaU0}) and dotted lines are quadratic fits to the Fermi liquid form Eq.~(\ref{Eq:FermiKappa}). The apparent kink-like behaviour 
for $U=2.5$ is due to the  sharp crossover between the Fermi liquid and the bad metal (local moments fixed point). Both $T$ and $U$ are measured in units of $W$.}
\label{Fig:CompressibilityVsTFnU}
\end{figure}
\end{center}

In Fig.~\ref{Fig:CompressibilityAndZT0.0} we also show the evolution of the zero temperature charge compressibility, $\kappa_{e}$, as a function of correlation strength, $U$. 
The charge compressibility continuously goes to zero with increasing, $U$. As mentioned earlier, $1/\kappa_{e}$ can be thought of as the energy required to 
add/remove an electron to/from the systems. Hence it gets increasingly harder to add or remove an electron into the system as we increase $U$, \textit{i.e.}, the system 
increasingly becomes incompressible and finally at $U=U_{c}$ the system becomes completely incompressible. Note that at $U=0$, $\kappa_{e}=N_{0}(E_{F})=2/\pi$, as it 
should be. A similar decrease in charge compressibility with increasing $U$ was observed in exact diagonalization calculations for the Hubbard model on the triangular 
lattice at half-filling~\cite{KokaljPRL2013}.        
% --------------------------------------------------------------------------------------------------------------------------------------------------------------
% --------------------------------------------------------------------------------------------------------------------------------------------------------------
In Fig.~\ref{Fig:CompressibilityVsTFnU} we show the temperature dependence of the charge compressibility for a range of values of $U$. In the non-interacting case ($U=0$) 
 we can  show that
\begin{eqnarray}
\kappa_{e}(T) = \int\limits_{-\infty}^{+\infty}\left(-\frac{\partial n_{F}(\omega)}{\partial \omega}\right) N_{0}(\omega)
\label{Eq:KappaU0}
\end{eqnarray}
This expression is similar to that for the Pauli spin susceptibility, $\chi(T)$, and the associated temperature dependence is shown in Figure \ref{Fig:CompressibilityVsTFnU}
as a dashed line. The steady decrease in the charge compressibility withincreasing temperature is largely due to the broadening of the Fermi-Dirac distribution function.

Using standard integral expressions involving the Fermi function~\cite{ashcroft1976solid}
\begin{eqnarray}
\int\limits_{-\infty}^{+\infty} H(\epsilon) n_{F}(\epsilon) d\epsilon = \int\limits_{-\infty}^{\mu} H(\epsilon) d\epsilon + \frac{\pi^{2}}{6}\left(k_{B}T\right)^{2} H'(\mu)+\cdots
\end{eqnarray}
the expression for the charge compressibility in Eq.~(\ref{Eq:KappaU0}) reduces to 
\begin{eqnarray}
\kappa_{e}(T)=N_{0}(\mu)+\frac{\pi^{2}}{6}\left(k_{B}T\right)^{2}\left . \frac{d^{2}N_{0}(\omega)}{d\omega^{2}}\right|_{\omega=\mu}+\cdots.
\end{eqnarray} 
For the Bethe lattice this reduces to
\begin{eqnarray}
\kappa_{e}(T)\simeq {2 \over \pi W}
\left[1-\frac{\pi^2}{6}\left(\frac{k_{B}T}{W}\right)^{2}+\cdots\right].
\label{Eq:FermiKappa0}
\end{eqnarray}
In the interacting case we expect in the Fermi liquid state 
\begin{eqnarray}
\kappa_{e}(T)\simeq \kappa_{e}(0)\left[1-\frac{\delta}{Z^{2}}\left(\frac{k_{B}T}{W}\right)^{2}+\cdots\right],
\label{Eq:FermiKappa}
\end{eqnarray}
with $\delta\sim 1$. So, because of the thermal broadening effects, just as in the case of the spin susceptibility, the charge compressibility in the Fermi liquid state 
will decrease quadratically in temperature at low temperatures. This quadratic dependence is shown in Fig. \ref{Fig:CompressibilityVsTFnU} as dotted lines that have been
fitted to the low temperature behaviour. This explains the rapid decrease of the charge compressibility at low temperatures in the coherent Fermi liquid state, because the 
temperature scale for the decrease is that of the coherence temperature which close to the Mott transition becomes very small \cite{MerinoPRB2000,GeorgesRMP1995}.  
% --------------------------------------------------------------------------------------------------------------------------------------------------------------
% --------------------------------------------------------------------------------------------------------------------------------------------------------------
\subsection{Resistivity}
\begin{center}
\begin{figure}[htbp]
\includegraphics[scale=0.36,clip=]{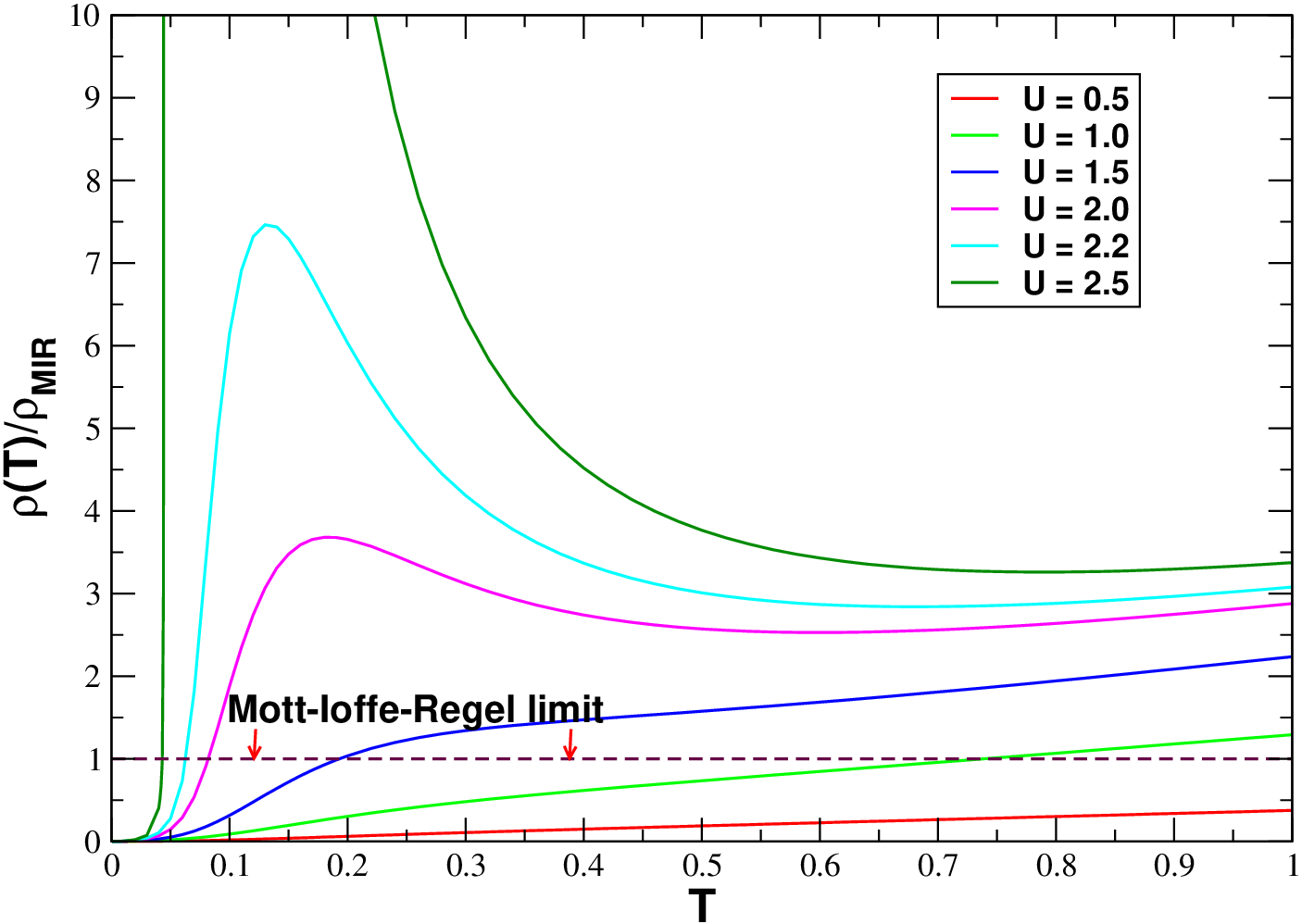}
\caption{(Color online) Resistivity shows violation of the Mott-Ioffe-Regel limit ($\rho_{MIR}=h a/e^{2}$) with increasing $U$. This is consistent with the 
picture that transport becomes increasingly incoherent with increasing correlation effects. Both $T$ and $U$ are measured in units of $W$.}
\label{Fig:ResistivityVsTFnU}
\end{figure}
\end{center}

In Fig.~\ref{Fig:ResistivityVsTFnU} we show the temperature dependence of the resistivity, $\rho(T)$, scaled by the Mott-Ioffe-Regel limit, $\rho_{MIR}=h a/e^{2}$, 
for various correlation strengths, $U$. In the weakly correlated regime ($U=0.5$) the resistivity is well within the Mott-Ioffe-Regel limit in the entire temperature 
range up to $W$. Hence, the transport can be characterised by weak scattering of coherent quasi-particle states. At very low temperatures ($T < T_{K} \ll W$) 
the resistivity is proportional to $T^{2}$ as expected in the coherent Fermi-liquid regime. The $T^{2}$ behaviour is due to the fact that in the Fermi liquid regime 
the imaginary part of the self-energy, or equivalently the inverse of quasi-particle life time ($\tau_{qp}^{-1}$) is  proportional to $T^{2}$ (or $\omega^{2}$ 
at $T=0$). At high temperatures ($T \gg T_{K}$), the resistivity is roughly
linear in $T$ and this corresponds to the incoherent (bad metal) regime of 
transport~\cite{DengPRL2013}. 

As we increase $U$, the resistivity smoothly crosses the Mott-Ioffe-Regel limit and in the strongly correlated regime ($U=2.0$ and above) the resistivity far exceeds the MIR 
limit. This is due to the sharp crossover from the strong coupling (Fermi liquid) regime to the local moment (bad metal) regime in the strong correlation regime and is 
consistent with the picture of fragile quasi-particles states in the strong correlation regime. 

In elemental crystals one can distinguish metals and insulators by the temperature dependence of the resistivity. It is monotonically increasing (decreasing) with increasing 
temperature for metals (insulators). However, this criteria is unreliable for strongly correlated electron materials. For example, a non-monotonic temperature dependence of 
the resistivity in a bad metal is seen experimentally in a number of organic charge salts. (See for example the inset of Figure 2(a) of Ref. \onlinecite{kurosaki}).
Thus it is important to note that even though the derivative $d\rho(T)/dT$ changes sign for some curves in Fig. \ref{Fig:ResistivityVsTFnU} there is no metal-insulator 
transition, i.e. all the curves are for the metallic phase. This is evident from the finite temperature spectral function, $A_{d}(\omega)$ at $\omega=0$ and the non-zero 
charge compressibility $\kappa_{e}$. 

It should also be stressed that for the given choice of $U$ we are still far away from Mott transition at $U_{c} \simeq 3.4$. This is also evident from the relatively large 
quasi-particle weight $Z\sim 0.2$ even for $U=2.5$ where the MIR limit is violated by a factor of 100. So, the transport in this bad metal phase is incoherent in nature.    
% --------------------------------------------------------------------------------------------------------------------------------------------------------------
% --------------------------------------------------------------------------------------------------------------------------------------------------------------
\subsection{Diffusivity}
\begin{center}
\begin{figure}[htbp]
\includegraphics[scale=0.36,clip=]{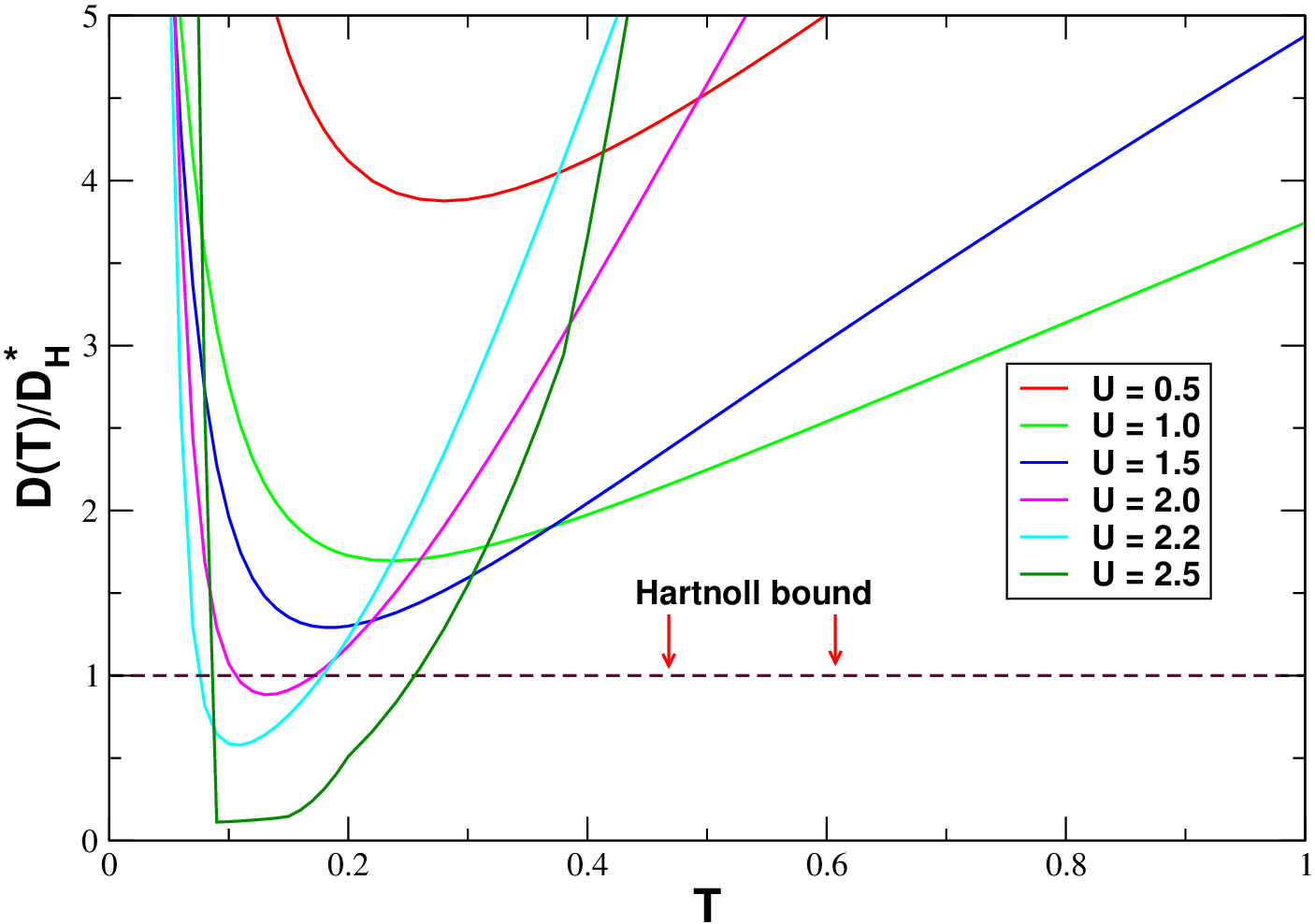}
\caption{(Color online) Scaled diffusivity shows violation of Hartnoll's bound in the strongly correlated incoherent regime of transport in the low temperature 
region. However, in strongly correlated systems at large temperatures ($T\gg T_{K}$) and for weakly correlated systems at all temperatures the bound is respected. Kink 
like behaviour for $U=2.5$ is due to sharp crossover between the Fermi liquid and the local moment regime. Both $T$ and $U$ are measured in units of $W$.}
\label{Fig:ScaledDiffusivityHartnoll}
\end{figure}
\end{center}

Finally, using the Nernst-Einstein relation we calculate the charge diffusivity. In Fig.~\ref{Fig:ScaledDiffusivityHartnoll} we show the scaled diffusivity, 
$D(T)/D_{H}^{*}$, as a function of temperature ($T$) for various correlation strengths ($U$). In the weakly correlated regime ($U=0.5$) and moderately 
correlated regime ($U=1.0$ and $U=1.5$) the scaled diffusivity satisfies Hartnoll's bound. However in the strongly correlated regime, $U=2.0$ and above, the scaled diffusivity 
shows violation of Hartnoll's bound in the low temperature region. But at high temperatures $T \gg T_{K}$ the scaled diffusivity satisfies Hartnoll's bound.
It is important to mention that the kink like behaviour at around $T \sim 0.05$ for $U=2.5$ is closely related to the 
sharp crossover between the Fermi liquid fixed point and the 
local moment fixed point in the effective single impurity Anderson model. As we can see from Fig.~\ref{Fig:ScaledDiffusivityHartnoll} the violation of Hartnoll's bound in 
strongly correlated systems is in the crossover region between coherent (Fermi liquid) regime and incoherent (local moment) regime. The magnitude of violation increases with 
increasing $U$ due to increased sharpness in crossover region. In the high temperature region, where the resistivity is roughly linear in $T$ the scaled diffusivity is well 
above the Hartnoll's bound, provided one uses the renormalised Fermi velocity. 
% --------------------------------------------------------------------------------------------------------------------------------------------------------------
% --------------------------------------------------------------------------------------------------------------------------------------------------------------

\begin{center}
\begin{figure}[htbp]
\includegraphics[scale=0.36,clip=]{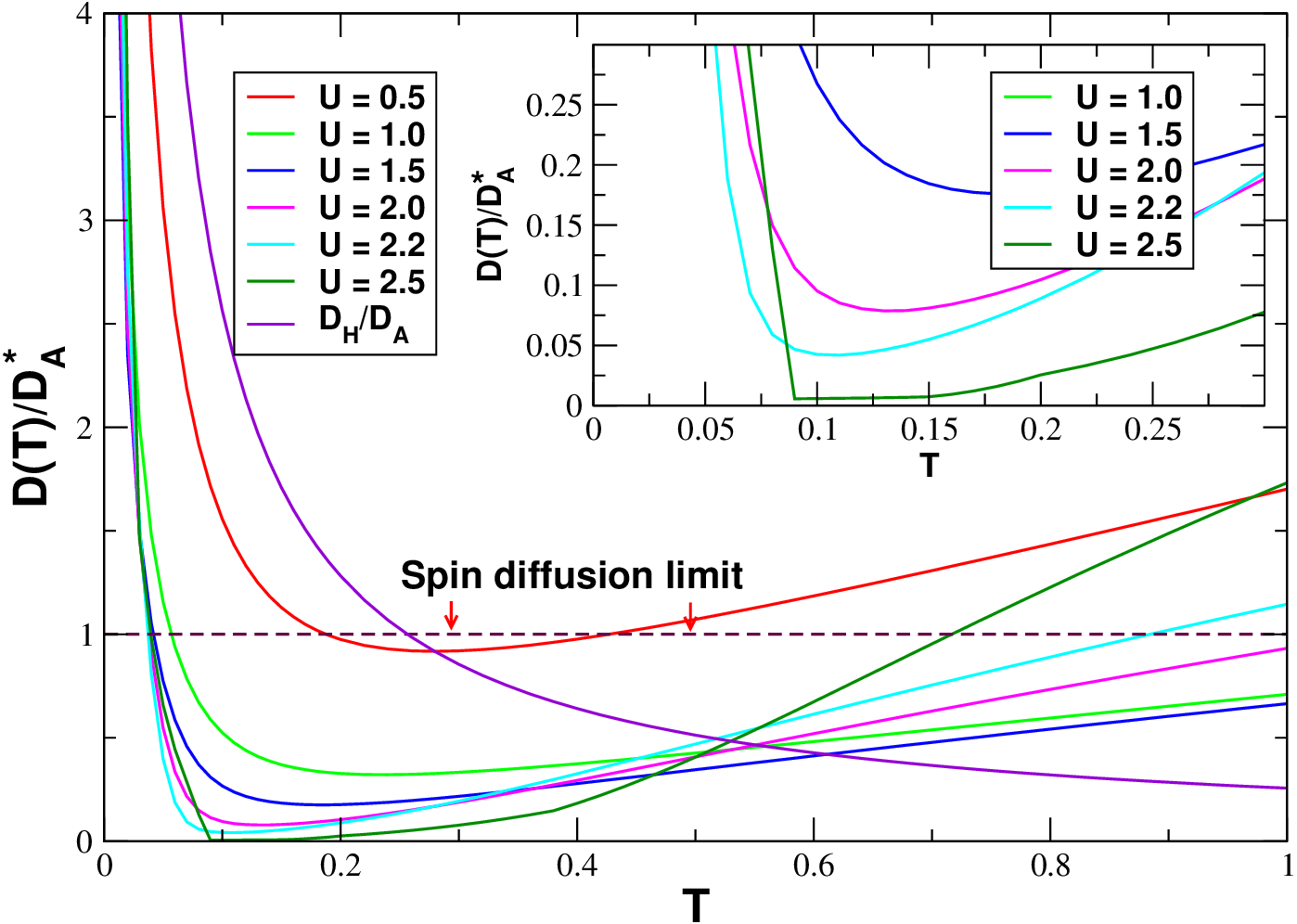}
\caption{(Color online) Scaled diffusion constant for charge transport also violates quantum limit for spin diffusion constant in the incoherent regime of 
transport. Inset : Detailed plot near the origin shows small yet non-vanishing scaled diffusivity for $U=2.5$. $D_{H}/D_{A}$ traces out the temperature 
dependence, $1/3\alpha k_{B}T$, of the Hartnoll bound showing how for $T < 0.3$ it is larger than the proposed bound on the spin diffusion constant for cold atoms. 
Both $T$ and $U$ are measured in units of $W$.}
\label{Fig:ScaledDiffusivityAtomic}
\end{figure}
\end{center}

Finally, we compare the charge diffusivity to the quantum limit of the spin diffusion constant, $D_{s} \simeq 1.3\hbar/m$, experimentally observed~\cite{SommerNature2011} and 
theoretically calculated~\cite{EnssPRL2012} in the degenerate Fermi gas in the unitary limit.  In Fig.~\ref{Fig:ScaledDiffusivityAtomic} we show the scaled diffusivity, 
$D(T)/D_{A}^{*}$, as a function of temperature for various $U$. The scaled diffusivity also violates the quantum limit of spin diffusion constant, $D_{s}\simeq 1.3\hbar/m$. 
The violation is severe in the strongly correlated regime. All the temperature dependence is ultimately due to inherent temperature dependence of self energy, $\Sigma(\omega)$. 

It is important to mention that spin diffusion in charge neutral systems like the degenerate Fermi gas in the unitary limit has no clear relation to charge diffusion in electron 
liquids. In charged systems there are dynamical screening effects while such screening effects are not present in neutral atomic gases such as the strongly interacting degenerate 
Fermi gas at Feshbach resonance. Most interestingly in a Mott insulator the charge diffusion constant is zero while the spin diffusion constant is finite. The differences 
illustrate different mechanisms for charge and spin transport in strongly correlated systems and one should not necessarily expect any simple relationship between the spin and 
charge diffusion constants.

For a degenerate non-interacting Fermi gas in three dimensions the charge diffusion constant is given by $D=\frac{1}{3}\frac{\hbar}{m}k_{F}\ell$, where $k_{F}$ is the 
Fermi wave vector and $\ell$ is the mean free path. In the weak scattering limit $k_{F}\ell \gg 1$ and $D \gg \frac{1}{3}\frac{\hbar}{m}$. So, just like the upper limit (MIR) of 
resistivity we can define lower limit for the charge diffusion constant $D_{lim}=\frac{1}{3}\frac{\hbar}{m}$. In the weakly interacting quasi-particle regime of transport the 
limit will be renormalized to $D_{lim}^{*}=\frac{1}{3}\frac{\hbar}{m^{*}}$. The quantum spin diffusion limit in degenerate Fermi gas will roughly correspond to $k_{F}\ell \sim 4$ 
for charge diffusion in a condensed matter system and the diffusion will correspond to transport through coherent quasi-particle states.          
% --------------------------------------------------------------------------------------------------------------------------------------------------------------
% --------------------------------------------------------------------------------------------------------------------------------------------------------------
\section{Conclusions}
We have studied the conductivity, charge compressibility, and charge diffusivity in a single band Hubbard model using single site dynamical mean field theory. The 
calculated resistivity far exceeds the MIR limit in the strong correlation regime. The transport in the weakly correlated region can be characterized by diffusive 
scattering of coherent quasi-particle states but in the strongly correlated bad metal state the transport is incoherent. The charge compressibility decreases with 
increasing $U$ which corresponds to the fact that in the correlated regime, the energy cost to create a charge fluctuation increases with increasing $U$. Then using the 
Nernst-Einstein relation we calculated the charge diffusivity in the system. In the weakly and moderately correlated systems the scaled diffusivity respects Hartnoll's bound 
at all temperatures. However, in the strongly correlated systems the bound is violated in the crossover region between the coherent Fermi liquid regime and incoherent local 
moment regime. In the high temperature region ($T \gg T_{K}$) particularly in the region where resistivity is roughly linear in $T$ the bound is found to be respected for all 
interaction strengths. 

We also compared the calculated charge diffusivity against the quantum limit of spin diffusion observed in the degenerate Fermi gas in the unitary 
limit. The calculated diffusivity strongly violates the quantum limit of spin diffusion in the incoherent regime. So, within the single site DMFT approximation 
we do not observe any quantum limit to charge diffusion in the strongly correlated incoherent regime.

Hartnoll's proposed bound is based on the AdS/CFT correspondence and various conservation laws in fluids. But within single site DMFT approximation there is energy 
conservation but no momentum conservation at a given site. 
On the other hand, the spatial fluctuations neglected in single site DMFT can be systematically incorporated through other approximations such the dynamical cluster approximation 
(DCA)~\cite{MaierRMP2005,LinPRB2010} in 
which momentum is conserved within the cluster, bath as well as at the boundary of the cluster should be able to address this issue. One might also consider how vertex 
corrections could modify the results. For a doped Hubbard model it was found in a 4 site DCA calculation that the vertex corrections to the optical conductivity were not 
significant, except very close to the Mott insulator~\cite{Lin}. A study of the doped two-dimensional Hubbard model using a two-particle self-consistent approach found that 
vertex corrections altered the calculated resistivity by less than a factor of two~\cite{BergeronPRB2011}.

As has been pointed out \cite{LiuPT2012} the interacting many electron system in strongly correlated materials so far has no gravity description and hence the strongly coupled 
gauge theory has no dual description. To be more precise, holographic quantum systems associated with known gravity descriptions have no direct relation with strongly correlated 
electron systems. Furthermore, the AdS/CFT correspondence only describes conformally invariant field theories and except for one dimensional systems at a quantum critical point, 
it is not clear that strongly correlated electron systems are conformally invariant~\cite{AndersonPT2013}.
             
\begin{acknowledgments}
We acknowledge helpful discussions with X. Deng, J. K. Freericks,
S. Hartnoll, H. R. Krishnamurthy, B. J. Powell, T. V. Ramakrishnan,  
D. Tanaskovic, A. Taraphder,   N. Vidhyadhiraja, and J. Vucicevic. This work 
was supported by an Australian Research Council Discovery Project grant.
\end{acknowledgments}
\vspace*{0.2cm}
\section*{APPENDIX: Derivation of the Nernst-Einstein relation}
Fick's law for diffusion is given by
\begin{eqnarray}
\mathbf{j}_{m}(\mathbf{r})=-D(T)\boldsymbol{\nabla} n(\mathbf{r})
\end{eqnarray}
where $D(T)$ is the diffusion constant, $\mathbf{j}_{m}(\mathbf{r})$ is the \textit{mass current} and $n(\mathbf{r})$ is the \textit{local particle number}.
On the other hand Ohm's law for electrical conductivity is given by
\begin{eqnarray}
\label{Eq:OhmsLaw}
\mathbf{j}_{e}(\mathbf{r})=\sigma(T) \mathbf{E}(\mathbf{r}) 
\end{eqnarray}
where $\sigma(T)$ is the electrical conductivity, $\mathbf{j}_{e}(\mathbf{r})$ is the \textit{electric current} and $\mathbf{E}(\mathbf{r})$ is the 
external electric field. We have
\begin{eqnarray}
\label{Eq:jeexpanded}
\mathbf{j}_{e}(\mathbf{r}) = e \mathbf{j}_{m}(\mathbf{r}) = -e D(T) \frac{\partial n}{\partial \mu} \boldsymbol{\nabla} \mu (\mathbf{r})
\end{eqnarray}
where $\mu (\mathbf{r})= \mu_{0}+e \phi(\mathbf{r})$ is the chemical potential in the presence of external field and $\mu_{0}$ is that in the absence of external field and 
$\phi(\mathbf{r})$ is the electric potential. 
Then
\begin{eqnarray} 
\label{Eq:Gradmu}
\boldsymbol{\nabla} \mu (\mathbf{r}) = e \boldsymbol{\nabla} \phi(\mathbf{r}) = -e \mathbf{E}(\mathbf{r}). 
\end{eqnarray}
Combining Eq.~(\ref{Eq:jeexpanded}) and~(\ref{Eq:Gradmu}) gives
\begin{eqnarray}
\mathbf{j}_{e}(\mathbf{r})=e^{2}\frac{\partial n}{\partial \mu} D(T) \mathbf{E}(\mathbf{r}). 
\end{eqnarray}
Comparing this with Ohm's law in Eq.~(\ref{Eq:OhmsLaw}) we finally get
\begin{eqnarray}
\sigma(T)=e^{2}\frac{\partial n}{\partial \mu} D(T).
\end{eqnarray}

\end{document}